\documentstyle[prd,preprint,aps]{revtex}
\begin{document}
\draft
\title{Higgs Pair-Production in the Standard Model at Next Generation
       Linear $e^+e^-$ Colliders}

\author{A. Guti\'errez-Rodr\'{\i}guez $^{1}$, M. A. Hern\'andez-Ru\'{\i}z $^{2}$
        and O. A. Sampayo $^{3}$}

\address{(1) Facultad de F\'{\i}sica, Universidad Aut\'onoma de Zacatecas\\
         Apartado Postal C-580, 98060 Zacatecas, Zacatecas M\'exico.}

\address{(2) Facultad de Ciencias Qu\'{\i}micas, Universidad Aut\'onoma de Zacatecas\\
         Apartado Postal 585, 98060 Zacatecas, Zacatecas M\'exico.}

\address{(3) Departamento de F\'{\i}sica, Universidad Nacional del Mar del Plata\\
         Funes 3350, (7600) Mar del Plata, Argentina.}
\date{\today}
\maketitle

\begin{abstract}
We study the Higgs pair-production in the Standard Model of the strong and
electroweak interactions at future $e^{+}e^{-}$ collider energies, with the
reaction $e^{+}e^{-}\rightarrow t \bar t HH$. We evaluated the total cross
section of $t\bar tHH$ and calculate the number total of events considering
the complete set of Feynman diagrams at tree-level. The numerical computation
is done for the energy which is expected to be available at a possible Next
Linear $e^{+}e^{-}$ Collider: with center-of-mass energy $800, 1600$ $GeV$ and
luminosity 1000 $fb^{-1}$.
\end{abstract}

\pacs{PACS: 13.85.Lg, 14.80.Bn,}


\section{Introduction}

In the Standard Model (SM) \cite{Weinberg} of particle physics, there are three
types of interactions of fundamental particles: gauge interactions, Yukawa
interactions and the Higgs boson self-interaction. The Higgs boson \cite{Higgs}
plays an important role in the SM; it is responsible for generating the masses
of all the elementary particles (leptons, quarks, and gauge bosons). However,
the Higgs-boson sector is the least tested one in the SM, in particular the
Higgs boson self-interaction. In the SM, the profile of the Higgs particle is
uniquely determined once its mass $M_H$ is fixed. The decay width, the branching
ratios and the production cross sections are given by the strength of the
Yukawa couplings to fermions and gauge bosons, the scale of which
is set by the masses of these particles. Unfortunately, the mass
Higgs boson is a free parameter.

The only available information on $M_H$ is the lower limit $M_H\geq 114.1$
$GeV$ established at LEP2 \cite{LEP}. The collaborations have also reported a
2.1$\sigma$ excess of events beyond the expected SM backgrounds consistent with
a SM like Higgs boson with a mass $M_H=115^{+1.3}_{-0.9}$ $GeV$ \cite {LEP}.
Furthermore, the accuracy of the electroweak data measured at LEP, SLC, and the
Tevatron provides sensitivity to $M_H$: the Higgs boson contributes
logarithmically, $\propto\log(\frac{M_H}{M_W})$, to the radiative corrections
to the $W/Z$ boson propagators. A recent analysis yields the value
$M_H=88^{+60}_{-37}$ $GeV$ corresponding to 95$\%$ C.L.

The search for Higgs boson is one of the main missions of present and future
high-energy colliders. The observation of this particle is of major
importance for the present understanding of the interactions of the
fundamental particles.

The trilinear Higgs self-coupling can be measured directly in
pair-production of Higgs particles at hadron and high-energy $e^+e^-$ linear
colliders. Higgs pairs can be produced through double Higgs-strahlung of $W$
or $Z$ bosons \cite{Gounaris,Ilyin,Djouadi,Kamoshita}, $WW$ or $ZZ$ fusion
\cite{Ilyin,Boudjema,Barger,Dobrovolskaya,Dicus}; moreover through
gluon-gluon fusion in $pp$ collisions \cite{Glover,Plehn,Dawson} and
high-energy $\gamma\gamma$ fusion \cite{Ilyin,Boudjema,Jikia} at photon
colliders. The two main processes at $e^+e^-$ colliders are double
Higgs-strahlung and $WW$ fusion:

\begin{eqnarray}
\mbox{double Higgs-strahlung}&:& e^+e^- \to ZHH  \nonumber \\
\mbox{$WW$ double-Higgs fusion}&:& e^+e^- \to \bar\nu_e \nu_e HH.
\end{eqnarray}

The $ZZ$ fusion process of Higgs pairs is suppressed by an order of magnitude
since the electron-Z coupling is small. However, the process $e^+e^-
\rightarrow t \bar t H, $ has been extensively studied. This three-body process
is important because it is sensitive to Yukawa couplings. The inclusion of
four-body processes with heavy fermions $f$, $e^{+}e^{-}\rightarrow f\bar f HH$
\cite{Ilyin} in which the SM Higgs boson is radiated by a $t(\bar t)$
quark, at future $e^{+}e^{-}$ colliders  \cite{NLC,NLC1,JLC} with a c.m. energy
in the range of 500 to 1600 $GeV$, such as the TESLA machine \cite{TESLA} is
necessary in order to know its impact on the three-body mode processes and also
to search for new relations that could have a clear signature of the Higgs
boson production.

Moreover, this process depends on the Higgs boson triple self-coupling, which
could lead us to obtain the first non-trivial information on the Higgs
potential. We are interested in finding regions that could allow the
observation of the process $t\bar tHH$ at the next generation of high energy
$e^{+}e^{-}$ linear colliders. We consider the complete set of Feynman diagrams
at tree-level (Fig.1) and used the CalcHep \cite{Pukhov} packages for the
evaluation of the amplitudes and of the cross section.

This paper is organized as follows: In Sec. II we present the total cross section
for the process $e^{+}e^{-}\rightarrow t \bar t HH$ at next generation linear
$e^{+}e^{-}$ colliders, and in Sec. III, we give our conclusions.

\section{Cross Section of the Higgs Pairs Production in the SM at Next
            Generation Linear Positron-Electron Colliders}

In this paper, we evaluate the total cross section of the Higgs pair-production
in the SM at next generation linear $e^{+}e^{-}$ colliders.

For the SM parameters, we have adopted the following: the angle of
Weinber $\sin^2\theta_W=0.232$, the mass ($m_t=175$ $GeV$) of the
top, the mass ($m_{Z^0}=91.2$ $GeV$) of the $Z^0$, with the mass
$M_H$ of the Higgs boson having been taken as inputs \cite{Review}.

We have considered the high energy stage of a possible Next Linear $e^{+}e^{-}$
Collider with $\sqrt{s}=800, 1600$ $GeV$ and design luminosity 1000 $fb^{-1}$.
In the evaluation of the amplitudes and of the cross section, we used the
CalcHep \cite{Pukhov} packages.

In order to illustrate our results of the production of Higgs pairs in the SM,
we present in Fig. 2 a plot for the total cross section as a function of Higgs
boson mass $M_{H}$. We observe in this figure that the total cross
section for the double Higgs production is of the order of $0.02$ $fb$
for Higgs masses in the lower part of the intermediate range. The cross
sections are at the level of a fraction of femtobarn, and they quickly drop as
they approach the kinematic limit. In these conditions, it would be very difficult
to extract any useful information about the Higgs self-coupling from the studied
process except that the $e^+e^-$ machine works with very high luminosity.

Fig. 3 shows the total cross section as a function of the center-of-mass
energy $\sqrt{s}$ for two representative values of the Higgs mass $M_H=110,
130$ $GeV$. We observe that the cross section is very sensitive to the Higgs
boson mass and decreases when $M_H$ increases. Our conclusion is that
for an intermediate Higgs boson, a visible number of events would be produced,
as is illustrated in Table I.

For center-of-mass energies of 800-1600 $GeV$ and high
luminosity, the possibility of observing the process $t\bar tHH$ is
promising as shown in Table I.

\vspace*{5mm}

\begin{center}
\begin{tabular}{|c|c|c|}
\hline Total Production of Higgs Pairs & \multicolumn{2}{c|}{$t\bar t HH$} \\
\cline{2-3} &  &  \\ \hline\hline $M_H(GeV)$ & $\sqrt{s}=800$ $GeV$ &
$\sqrt{s}=1600$ $GeV$ \\ \hline
 110 & 14 & 19 \\ 130 & 6 & 15 \\ 150 & 2 & 11 \\
170 & 1 & 9 \\ 190 &  & 7 \\ \hline
\end{tabular}
\end{center}

\begin{center}
Table I. Total production of Higgs pairs in the SM for ${\cal L}=1000$ $fb^{-1}$
and $m_t=175$ $GeV$.
\end{center}

We also include in Fig. 4 a contours plot for the number of events of the
studied process, as a function of $M_H$ and $\sqrt{s}$.

\section{Conclusions}

In conclusion, the double Higgs production in association with
$t(\bar t)$ quarks ($e^{+}e^{-}\rightarrow t \bar t HH$) will be
observable at the Next Generation Linear $e^+e^-$ Colliders. The
study of this process is important in order to know their impact on the
3-body process and it could be useful to probe anomalous $HHH$
coupling given the following conditions: very high luminosity,
center-of-mass large energy and intermediate range Higgs mass.

\hspace{3cm}

\begin{center}
{\bf Acknowledgments}
\end{center}

This work was supported in part by {\it Consejo Nacional de Ciencia y
Tecnolog\'{\i}a} (CONACyT) (Proyect: 40729-F), {\it Sistema Nacional de Investigadores} (SNI)
(M\'exico) and Programa de Mejoramiento al Profesorado (PROMEP). O.A. Sampayo
would like to thank CONICET (Argentina). The authors would also like to
thank Maureen Sophia Harkins Kenning for revising the manuscript.

\newpage

\begin{center}
{\bf FIGURE CAPTIONS}
\end{center}

\vspace{5mm}

\bigskip

\noindent {\bf Fig. 1} Feynman Diagrams at tree-level for $e^{+}e^{-}
\rightarrow t\bar t HH$.

\bigskip

\noindent {\bf Fig. 2} Total cross section of the Higgs pairs production
$e^{+}e^{-}\rightarrow t\bar t HH$ as function of the Higgs mass $M_H$ for
$\sqrt{s} =800, 1600$ $GeV$ with $m_{t} = 175$ $GeV$.

\bigskip

\noindent {\bf Fig. 3} Total cross section of the Higgs pairs
production $e^{+}e^{-}\rightarrow t\bar t HH$ as function of the
center-of-mass energy $\sqrt{s}$ for two representative values
of the Higgs mass $M_H=110, 130$ $GeV$ with $m_t=175$ $GeV$.

\noindent {\bf Fig. 4} Contours plot for the number of events as a function of
$M_H$ and $\sqrt{s}$.


\newpage

\begin{references}

\bibitem{Weinberg} S. Weinberg, Phys. Rev. Lett. {\bf 19}, (1967) 1264; A.
                   Salam, in {\it Elementary Particle Theory}, ed. N. Southolm (Almquist and
                   Wiksell, Stockholm, 1968), p.367; S.L. Glashow, Nucl. Phys. {\bf 22}, (1967)257.

\bibitem{Higgs} P. W. Higgs, Phys. Lett. {\bf 12}, (1964) 132; P. W. Higgs,
                Phys. Rev. Lett. {\bf 13}, (1964) 508; P. W. Higgs, Phys. Rev. Lett. {\bf 145},
                (1966) 1156; F. Englert, R. Brout, Phys. Rev. Lett. {\bf 13}, (1964) 321;
                G. S. Guralnik, C. S. Hagen, T. W. B. Kibble, Phys. Rev. Lett. {\bf 13}
                (1964), 585.

\bibitem{LEP} The LEP Higgs Working Group, hep-ex/0107029 and
              hep-ex/0107030.


\bibitem{Gounaris} G. Gounaris, D. Schildknecht and F. Renard, Phys. Lett.
                   {\bf B83}, (1979) 191 and (E) {\bf 89B}, (1980) 437; V. Barger, T. Han and
                   R.J.N. Phillips, Phys. Rev. {\bf D38}, (1988) 2766.

\bibitem{Ilyin} V. A. Ilyin, A. E. Pukhov, Y. Kurihara, Y. Shimizu and T.
                Kaneko, Phys. Rev. {\bf D54}, (1996) 6717.

\bibitem{Djouadi} A. Djouadi, H. E. Haber and P. M. Zerwas, Phys. Lett.
                  {\bf B375}, (1996) 203; A. Djouadi, W. Kilian, M. M. Muhlleitner and P. M.
                  Zerwas, Eur. Phys. J. {\bf C10}, (1999) 27; P. Oslan, P. N. Pandita, Phys.
                  Rev. {\bf D59}, (1999) 055013; F. Boudjema and A. Semenov, hep-ph/0201219;
                  A. Djouadi, hep-ph/0205248.

\bibitem{Kamoshita} J. Kamoshita, Y. Okada. M. Tanaka and I. Watanabe,
                    hep-ph/9602224; D. J. Miller and S. Moretti, hep-ph/0001194; D. J. Miller
                    and S. Moretti, Eur. Phys. J. {\bf C13}, (2000) 459.

\bibitem{Boudjema} F. Boudjema and E. Chopin, Z. Phys. {\bf C73}, (1996) 85.

\bibitem{Barger} V. Barger and T. Han, Mod. Phys. Lett. {\bf A5}, (1990) 667.

\bibitem{Dobrovolskaya} A. Dobrovolskaya and V. Novikov, Z. Phys. {\bf C52}, (1991) 427.

\bibitem{Dicus} D. A. Dicus, K. J. Kallianpur and S. S. D. Willenbrock,
                Phys. Lett. {B200}, (1988) 187;  A. Abbasabadi, W. W. Repko, D. A. Dicus and
                R. Vega, Phys. Rev. {\bf D38}, (1988) 2770;  Phys. Lett. {\bf B213}, (1988) 386.

\bibitem{Glover} E. W. N. Glover and J. J. van der Bij, Nucl. Phys. {\bf B309}, (1988) 282.

\bibitem{Plehn} T. Plehn, M. Spira and P. M. Zerwas, Nucl. Phys. {\bf B479},
                (1996) 46;  (E) Nucl. Phys. {\bf B531}, (1998) 655.

\bibitem{Dawson} S. Dawson, S. Dittmaier and M. Spira, Phys. Rev. {\bf D58},
                 (1998) 115012.

\bibitem{Jikia}  G. Jikia, Nucl. Phys. {\bf B412}, (1994) 57.

\bibitem{NLC} NLC ZDR Desing Group and the NLC Physics Working Group, S.
              Kuhlman {\it et al.}, {\it Physics and Technology of the Next Linear Collider},
              hep-ex/9605011.

\bibitem{NLC1} The NLC Design Group, C. Adolphsen {\it et al.} Zeroth-Order
               Design Report  for the {\it Next Linear Collider}, LBNL-PUB-5424, SLAC
               Report No. 474,  UCRL-ID-124161 (1996).

\bibitem{JLC} JLC Group, JLC-I, KEK Report No. 92-16, Tsukuba (1992).

\bibitem{TESLA} TESLA Technical Desing Report, Part III, DESY-01-011C, hep-ph/0106315.

\bibitem{Pukhov} {\it CompHEP - a package for evaluation of Feynman diagrams and
               integration over multi-particle phase space.}A.Pukhov et al,
               Preprint INP MSU 98-41/542, hep-ph/9908288

\bibitem{Review} Review of Particle Physics, Particle Data Group, Phys. Rev. {\bf D66}, (2002) 1.

\end{references}
\end{document}